\documentclass[english,prl, twocolumn, notitlepage, longbibliography, superscriptaddress]{revtex4-1}
\usepackage[T1]{fontenc}
\usepackage[latin9]{inputenc}
\usepackage{array}
\usepackage{verbatim}
\usepackage{amsmath}
\usepackage{amssymb}
\usepackage{graphicx}

\makeatletter

\providecommand{\tabularnewline}{\\}

\usepackage{amsmath}
\usepackage[colorlinks=true,linkcolor=MidnightBlue,urlcolor=MidnightBlue,citecolor=MidnightBlue,anchorcolor=MidnightBlue]{hyperref}\usepackage[dvipsnames]{xcolor}

\makeatother

\usepackage{babel}
\begin{document}
\title{Self-propulsion of active droplets without liquid-crystalline order}
\author{Rajesh Singh}
\email{rs2004@cam.ac.uk}
\affiliation{DAMTP, Centre for Mathematical Sciences, University of Cambridge,
Wilberforce Road, Cambridge CB3 0WA, United Kingdom}
\author{Elsen Tjhung}
\affiliation{DAMTP, Centre for Mathematical Sciences, University of Cambridge,
Wilberforce Road, Cambridge CB3 0WA, United Kingdom}
\affiliation{Department of Physics, University of Durham, Science Laboratories,
South Road, Durham DH1 3LE, UK}
\author{Michael E. Cates}
\affiliation{DAMTP, Centre for Mathematical Sciences, University of Cambridge,
Wilberforce Road, Cambridge CB3 0WA, United Kingdom}
\begin{abstract}
The swimming of cells, far from any boundary, can arise
in the absence of long-range liquid-crystalline order within the cytoplasm, but simple models of this effect are lacking. Here we present a two-dimensional model of droplet
self-propulsion involving two
scalar fields, representing the cytoplasm and a contractile cortex. An active stress results from coupling between these fields; self-propulsion results when rotational symmetry is spontaneously broken. The swimming speed is predicted, and shown numerically, to vary linearly with the
activity parameter and with the droplet area fraction. The model exhibits a Crowley-like
instability for an array of active droplets. 
\end{abstract}
\maketitle
Active fluids are an emerging class of nonequilibrium systems, where
energy is injected into the system \emph{locally} and continuously,
by the constituent particles themselves~\cite{Marchetti_2013}. Many
examples of active fluids are biological in nature, for example, actomyosin
networks inside the cell cytoskeleton~\cite{Svitkina_1997,Hawkins_2011,Poincloux_2011}
and dense suspensions of microtubules and kinesins \emph{in vitro}~\cite{Sanchez_2012,Guillamat_2018}.
In the case of actomyosin networks, each myosin motor can attach (and
detach) to two actin filaments and pull the two filaments inwards,
causing a net local contractile stress, whch drives the system out-of
equilibrium. In many cases, this local energy injection at the filament
scale can be translated into a macroscopic motion. For example, actomyosin
contraction at the rear of the cell cortex has been shown to play
an important role in the swimming motility of cells in a bulk fluid environment~\cite{Hawkins_2011,Poincloux_2011},
far from any boundary at which crawling can instead occur.

At the level of phase-field modeling and simulations, cell motility
is often described as the spontaneous motion of a droplet of active fluid~\cite{Tjhung_2015,Ziebert_2016,Camley_2017,Loisy_2020}.
The current field-theoretic understanding of cell swimming involves a scalar field $\phi$, coupled to polar or nematic liquid-crystalline
order, described by a vector or a second-rank tensor~\cite{Tjhung_2012,Tjhung_2015,Ziebert_2016}.
The scalar field delineates the cell's interior ($\phi>0$) from its exterior ($\phi<0$) whereas
the vectorial/tensorial field describes bulk internal alignment of a uniform or cortical `cytoskeleton'.
 The propulsion mechanism then relies on
a \emph{discrete} broken symmetry along a pre-existing axis of orientational
order, such as a spontaneous splay transition~\cite{Tjhung_2012}, or self-advection caused by net polymerization at the leading end of each polar filament~\cite{Tjhung_2015,Aranson_2016,Loisy_2020}. Thus, the vectorial/tensorial nature
of the order parameter is crucial to obtain self-propulsion in such
theories. On the other hand, experimental observations of cell swimming
suggests direct rotational symmetry breaking of the
actomyosin concentration delineating the cell cortex~\cite{Hawkins_2011,Poincloux_2011}, implying that liquid-crystalline order is not a pre-requisite
for self-propulsion in cells.

\begin{figure}[t]
\includegraphics[width=0.45\textwidth]{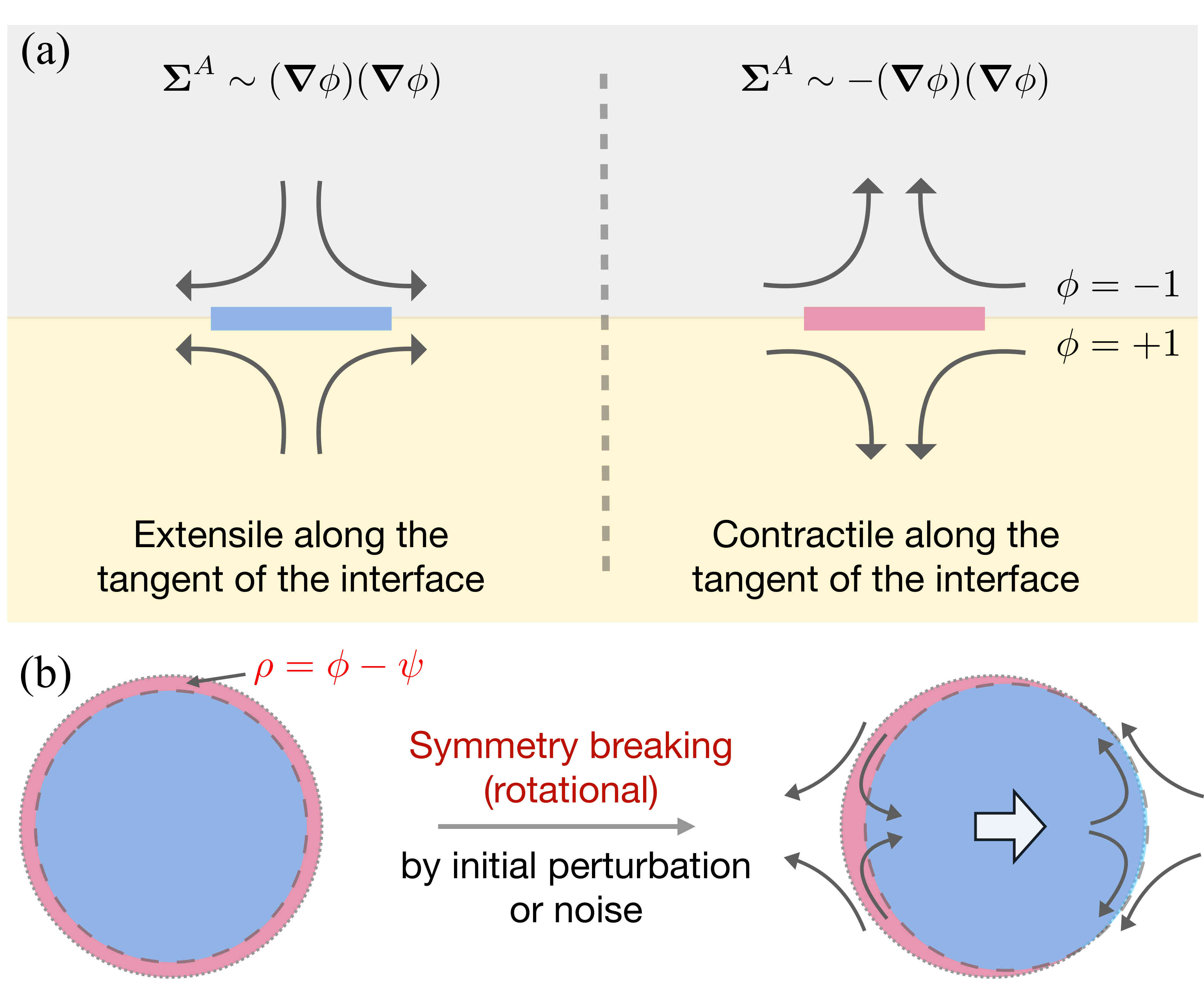} \caption{
(a) Flow driven by an active stress $\boldsymbol{\Sigma}^{A}$ on the
interface of the droplet phase field $\phi$. A positive (negative) $\boldsymbol{\Sigma}^{A}$
implies an extensile (contractile)
\textcolor{black}{ - shown by a blue (red) patch -}
fluid flow tangential
to the $\phi$-interface. 
(b) A cell is represented by $\phi$-droplet,
while the concentration of actomyosin in the cell cortex is given by
$\phi\mathcal{-}\psi$ (red ring). Under a small random initial
perturbation, the $\psi>0$ domain is displaced to the front
(right), creating excess
contractile stress at the back. This leads to an active flow 
which locks the $\psi$-droplet in the front, causing sustained 
self-propulsion along a direction selected by spontaneous symmetry breaking. \label{fig:1}}
\end{figure}

In this Rapid Communication, we present a field theory of self-propelling active
droplets, such as cells, in the absence of liquid-crystalline
order. Our theory is given in terms of two scalar fields $\phi$ and
$\psi$, which follow symmetry arguments of the Landau theory \cite{chaikin2000principles}, coupled
by the active stress in the momentum conservation equation. The $\phi$-field delineates
the interior/exterior of the droplet.  
The $\psi$ field is \textcolor{black}{an auxilary field which is related to a conserved scalar describing the local amount of active material ($\rho$). It is 
defined via $\rho = \phi-\psi$}. The arrangement of interest is where a cortex of nonzero $\rho$ [see the red region in Fig.\ref{fig:1}(b)] resides in the outer part of an otherwise passive droplet with constant overall mass density. This arrangement corresponds to a droplet with $\psi>0$ [see the blue core in Fig.\ref{fig:1}(b)] superimposed on a larger droplet of $\phi>0$. The difference of the two radii is then the thickness of a cortex 
surrounding the droplet, rendered contractile by, e.g., the action of myosin
motors.

Self-propulsion of the
active droplet emerges generically by the spontaneous
breakdown of rotational symmetry in the $\rho$ field. Once
this is broken, the droplet moves with a speed
that increases linearly with the activity strength and decreases
linearly with the droplet area fraction within a periodic domain. (We rationalize both scalings analytically.) We then extend our theory to
study an array of active droplets and identify an active-Crowley-like
instability. 

\emph{Model:} First, with two conserved scalar fields $\phi(\boldsymbol{r},t)$
and $\psi(\boldsymbol{r},t)$, we choose the free energy functional $\mathcal{F}[\phi,\psi]$
as: 
\begin{align}
\mathcal{F}[\phi,\psi] & =\int\Big(\frac{a}{2}\phi^{2}+\frac{b}{4}\phi^{4}+\frac{a'}{2}\psi^{2}+\frac{b'}{4}\psi^{4}-\beta\phi\psi\nonumber \\
 & +\frac{\kappa}{2}|\boldsymbol{\nabla}\phi|^{2}+\frac{\kappa'}{2}|\boldsymbol{\nabla}\psi|^{2}\Big)\,d\boldsymbol{r},\label{eq:F}
\end{align}
where $a=-b<0$, $a'=-b'<0$ and $\kappa,\kappa',\beta>0$.
\textcolor{black}{Although we choose $a=a'$ and $b=b'$, our results are robust against 
changing these parameters as long as the free energy admits the solution for a stable droplet. 
Connections of these free energy parameters to physical parameters - surface tension, interfacial width, thickness of cortex, etc - are given in Table 1 of the SI.}

Eq.\,(\ref{eq:F}) is adopted as the simplest way to stabilize two concentric phase-field domains (in $\phi$ and $\psi$) of unequal size. On identification of $\rho = \phi-\psi$, this becomes a droplet surrounded by a cortex, as required. 
Parameters are chosen so that each phase field approaches $\pm 1$ in the interior/exterior bulk phases. 
The $\beta>0$ term is an energetic coupling which favors maximizing the overlap of $\psi$ and $\phi$ fields.
Since both fields are conserved, the volumes of these $\phi,\psi$ droplets (equivalently, the $\phi$ droplet and its $\phi-\psi$ cortex) are separately constant in time.
We choose the initial volume of the $\phi$-droplet to be bigger than of the $\psi$-droplet, and thus,  the $\psi$-droplet
resides within the $\phi$-droplet giving a cortex in between  [see Fig.~\ref{fig:1}(b)].
The $\phi$-droplet then has
interfacial tension $\gamma_{0}\simeq\sqrt{-8\kappa a^{3}/9b^{2}}$~\cite{Cates_2017}, with similar expressions for the $\psi$-droplet. These tensions govern respectively the cortex/exterior and cortex/interior interfaces. 
Note that the $\beta$-term also renormalizes the interfacial 
tension $\gamma_{0}$; however, we choose $\beta/a=\beta/a'=0.01$, so that this difference is not appreciable. 

\begin{figure}[t]
\includegraphics[width=0.86\columnwidth]{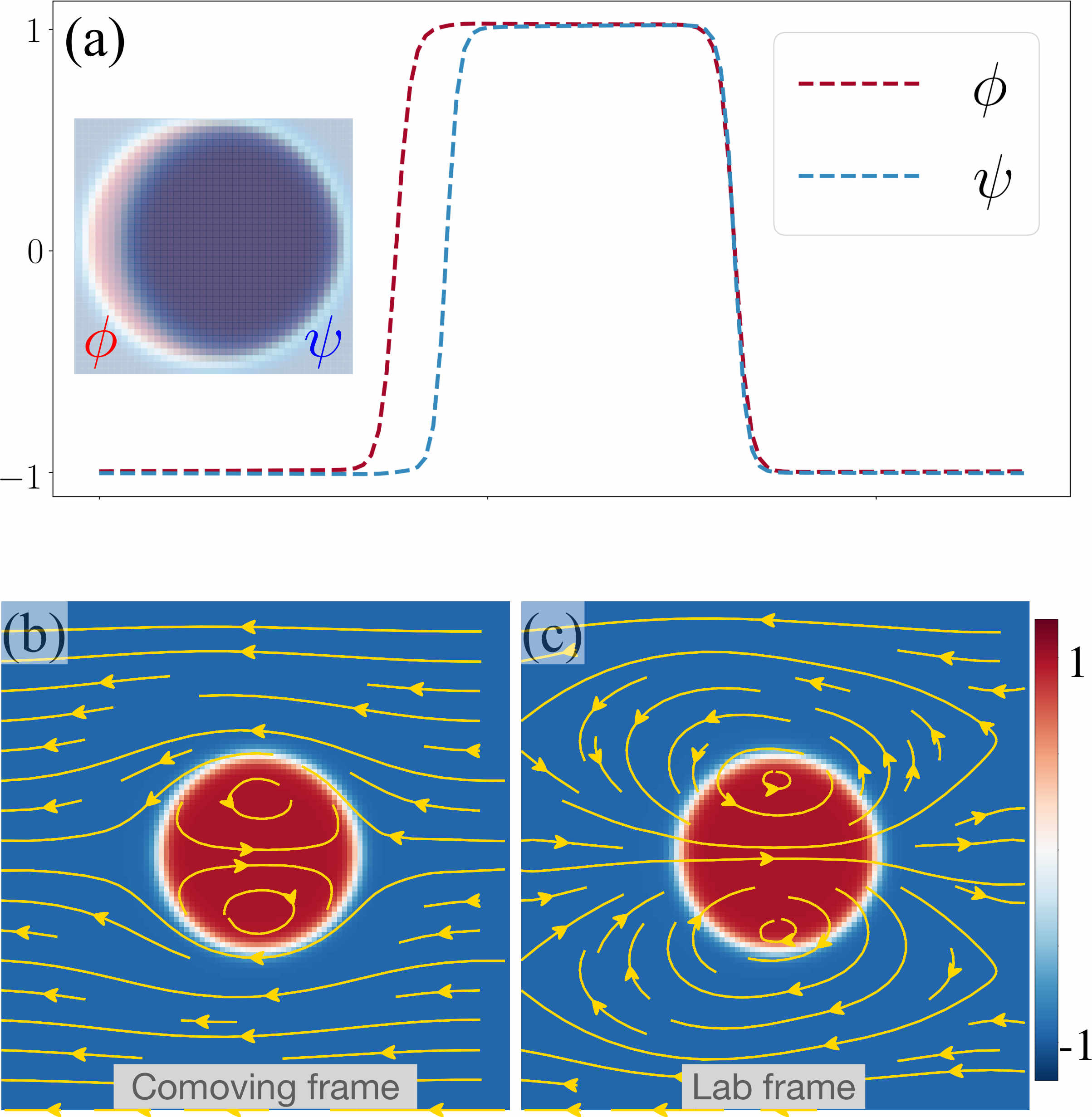} \caption{(a) \emph{Inset:} A snapshot of an active droplet steadily moving
in the $x$-direction. \emph{Main:} Corresponding plots of $\phi$
and $\psi$ as a function of $x$ at the cross-section passing through
the centre of the droplet. (b) and (c) show the streamlines of the fluid
flow of the same active droplet in (b) the comoving frame and (c) laboratory
frame. The streamlines are superimposed on the pseudo-color plot of
the $\phi$-field. 
\label{fig:2}}
\end{figure}

The only active term in our model is a contractile stress, which lives, for simplicity,
at the outer interface where $\phi$ passes through zero. This active stress is, however, modulated by the local concentration of cortical material which (in some units) is $\rho(\boldsymbol{r},t) = \phi-\psi \simeq -\psi >0$ [see Fig.~\ref{fig:2}(a)]. If rotational symmetry is maintained, the cortex is concentric with the droplet and the active stress is likewise symmetric (Fig.~\ref{fig:1}(b), left).
In the broken symmetry state, with more cortical material at the rear of the droplet, the active stress is larger there, creating a fluid flow that sustains the broken symmetry, by sweeping the cortex of actomyosin towards the rear so that $-\psi$ is larger there than at the front [Fig.~\ref{fig:1}(b) right].

This flow is governed by the hydrodynamic velocity $\boldsymbol{v}(\boldsymbol{r},t)$,
which describes the average velocity of the cellular materials plus
the solvent.
The conserved dynamics of $\phi$ and $\psi$ is then as follows:\begin{subequations}
\begin{align}
\frac{\partial\phi}{\partial t}+\boldsymbol{\nabla}\cdot\left(\phi\boldsymbol{v}-M^{\phi}\boldsymbol{\nabla}\frac{\delta\mathcal{F}}{\delta\phi}\right) & =0,\label{eq:phidot}\\
\frac{\partial\psi}{\partial t}+\boldsymbol{\nabla}\cdot\left(\psi\boldsymbol{v}-M^{\psi}\boldsymbol{\nabla}\frac{\delta\mathcal{F}}{\delta\psi}\right) & =0,\label{eq:psidot}
\end{align}
\end{subequations}such that the total $\int\phi\,d\boldsymbol{r}$
and $\int\psi\,d\boldsymbol{r}$ are constant in time. The first term
inside the parentheses describes advection of $\phi$ and $\psi$ by
the fluid velocity $\boldsymbol{v}$. The second term describes diffusion of $\phi$ and $\psi$ along the negative gradient
of the chemical potential $\delta\mathcal{F}/\delta\phi$ and $\delta\mathcal{F}/\delta\psi$,
respectively. $M^{\phi,\psi}>0$ are the mobilities for each field. 

In the limit of low Reynolds number, which is appropriate for sub-cellular
materials, the fluid flow $\boldsymbol{v}(\boldsymbol{r},t)$ is obtained
from solving the Stokes equation: 
\begin{equation}
\boldsymbol{\nabla}\cdot(\boldsymbol{\sigma}+\boldsymbol{\Sigma}^{E}+\boldsymbol{\Sigma}^{A})=\mathbf 0,\label{eq:Stokes}
\end{equation}
where $\boldsymbol{\sigma}=-p\boldsymbol{I}+\eta[(\boldsymbol{\nabla}\boldsymbol{v})+(\boldsymbol{\nabla}\boldsymbol{v})^{T}]$
is the Cauchy stress tensor in a fluid of viscosity $\eta$, $\boldsymbol{I}$
is the identity matrix, and $p$ is the isotropic pressure, which enforces
the incompressibility condition $\boldsymbol{\nabla}\cdot\boldsymbol{v}=0$.
$\boldsymbol{\Sigma}^{E}$ in (\ref{eq:Stokes}) is the equilibrium
interfacial stress, which is derived from the free energy functional
(\ref{eq:F})~\cite{Cates_2017}: 
\begin{equation}
\boldsymbol{\Sigma}^{E}=-\kappa(\boldsymbol{\nabla}\phi)(\boldsymbol{\nabla}\phi)-\kappa'(\boldsymbol{\nabla}\psi)(\boldsymbol{\nabla}\psi).\label{eq:SigmaE}
\end{equation}
Physically, $\boldsymbol{\Sigma}^{E}$ is the elastic response to
a deformation in the interface of the $\phi$- and $\psi$-droplet. 

$\boldsymbol{\Sigma}^{A}$
in (\ref{eq:Stokes}) is the active stress, which drives the system
out of equilibrium. The form of $\boldsymbol{\Sigma}^{A}$ is adapted
from the active model H~\cite{Tiribocchi_2015,Singh_2019}: 
\begin{equation}
\boldsymbol{\Sigma}^{A}=\alpha\psi(\boldsymbol{\nabla}\phi)(\boldsymbol{\nabla}\phi),\label{eq:SigmaA}
\end{equation}
where $\alpha\ge0$ is a cortical contractile activity parameter, such that the equilibrium
limit is recovered when $\alpha\rightarrow0$. From (\ref{eq:SigmaA}),
the activity is always localized at the interface of the $\phi$-droplet.
This differs from other models of active fluid droplets such as active nematics~\cite{Tjhung_2012,Blow_2014,Guillamat_2018,Giomi_2014},
where the active stress affects the bulk of the interior. The physical
significance of $\boldsymbol{\Sigma}^{A}$ is illustrated in Fig.~\ref{fig:1}(a).
Consider \textcolor{black}{a patch of active region on an interface separating $\phi=+1$ and $\phi=-1$ regions 
(see Fig.\ref{fig:1}a)}.
If $\alpha\psi>0$ so that $\boldsymbol{\Sigma}^{A}\propto(\nabla\phi)(\nabla\phi)$, the
active stress creates an extensile fluid flow in the tangential
direction of the interface. On the other hand if $\alpha\psi<0$, as will hold here, $\boldsymbol{\Sigma}^{A}\propto-(\nabla\phi)(\nabla\phi)$, and
the active stress creates a contractile flow tangential to
the interface, corresponding to actomyosin contractility $\alpha>0$
in the cell cortex. (Recall that the density of this cortex is $\rho\sim -\psi>0$ at the outer droplet interface.)
Note that in the
literature on active phase separation~\cite{Tiribocchi_2015,Singh_2019}, contractile and extensile
are defined with respect to the microswimmer orientation, normal to the interface.
The opposite convention, chosen here, refers to the tangential cortex layer and is used in the cellular 
literature~\cite{Hawkins_2011,Poincloux_2011}. 

Our numerical system consists of a two-dimensional (2D) square box with linear size
$L$ and periodic boundary conditions. We initialize a
$\phi$-droplet with radius $R$ at the centre of the box, and similarly
a $\psi$-droplet with a smaller radius $0.9R$; see Fig.~\ref{fig:1}(b) left. We then solve Eqns.~(\ref{eq:phidot}-\ref{eq:Stokes})
numerically using a pseudo-spectral method, as detailed in the SI \cite{Appendix}.

\emph{Mechanism of self-propulsion:} Now we will illustrate the mechanism
for self-propulsion. We fix the activity $\alpha$ to be finite and
positive. First, let us consider what happens when both $\phi$- and
$\psi$-droplets are concentric as shown in Fig.~\ref{fig:1}(b) left.
At the interface of the $\phi$-droplet, the value of $\psi$ is negative
and the same everywhere along the interface ($\psi\simeq-1$). Thus,
the active stress $\boldsymbol{\Sigma}^{A}$ (\ref{eq:SigmaA}) is
contractile and its magnitude is the same everywhere along the interface.
This represents an isotropic distribution of active cortex material, and thus, by symmetry, we should not expect to
see any motion.

Now imagine that we give a small displacement (which can come either
from thermal fluctuations or an initial perturbation -- we consider
the latter case here) to the $\psi$-droplet as shown in Fig.~\ref{fig:1}(b)
right. The essence of the symmetry breaking mechanism can be understood by considering 
a small initial displacement of the smaller droplet $\psi$ 
such that the interface of the $\phi$ droplet touches that of the $\psi$ droplet at one point 
(found below to be the `front' of the droplet when in motion) so that the $\rho$ cortex vanishes in thickness at this point.
Therefore at the front, the active stress $\boldsymbol{\Sigma}^{A}$
(\ref{eq:SigmaA}) is approximately zero, whereas at the back, the
active stress is finite and contractile. This excess contractile stress
at the back pulls the fluid from the front to the back along the interface
of $\phi$, as indicated by black arrows in Fig.~\ref{fig:1}(b)
right. This translates into persistent motion. Fig.~\ref{fig:2}(a)
shows the values of $\phi$ and $\psi$ measured at the cross-section,
which supports this mechanism. Physically, we have an accumulation of
actomyosin at the back of the cell which creates excess contractility
at the rear cell cortex. This is consistent with experiments on
cellular swimming motility in the absence of any boundary on which to crawl~\cite{Poincloux_2011,Ruprecht_2015}.

Fig.~\ref{fig:2}(b,c) show the steady state fluid velocity from
the full hydrodynamic simulations
in the co-moving (b) and laboratory frame (c). This fluid flow is as schematically shown in Fig.~\ref{fig:1}(b) right. Inside
the $\phi$-droplet, the differential active stress at the interface
generates a pair of counter-rotating vortices, which then squash the
$\psi$-droplet further to the front, giving a positive feedback.

Incidentally, this type of fluid flow is generic to all neutral (pure
quadrupolar flow \cite{kim2005}) squirmer droplets, whose motion is typically
driven by Marangoni flow along the interface~\cite{Thutupalli_2011,Jin_2017,Izzet_2019}.
From Figs.~\ref{fig:1}(b) and \ref{fig:2}(a), in the front hemisphere,
$\psi = \psi_f\simeq0$, whereas in the back hemisphere, $\psi = \psi_b\simeq-1$. The
active stress $\boldsymbol{\Sigma}^{A}$ renormalizes the surface
tension into an effective surface tension, $\gamma_{0}\rightarrow\gamma=\gamma_{0}(1-\alpha\psi/\kappa)$
\cite{Tiribocchi_2015,Singh_2019}, where $\gamma_{0}=\sqrt{-8\kappa a^{3}/9b^{2}}$
\cite{Cates_2017} is the surface tension without activity,
$\alpha=0$. This has the effect of a net change in surface tension
between the front and rear of the $\phi$-droplet, which is $\Delta\gamma=\alpha\Delta\psi\gamma_{0}/\kappa$,
where $\Delta\psi = \psi_f-\psi_b$. Thus, we have a Marangoni flow from low effective
surface tension (front) to high effective surface tension (back) as
shown in Fig. \ref{fig:2}(b), while Fig. \ref{fig:2}(c) contains
the same flow in the laboratory frame. The corresponding theoretical flow
around an active droplet, in an infinite 2D domain, is given in the first
figure of \cite{Appendix}.
\begin{figure}[t]
\includegraphics[width=0.95\columnwidth]{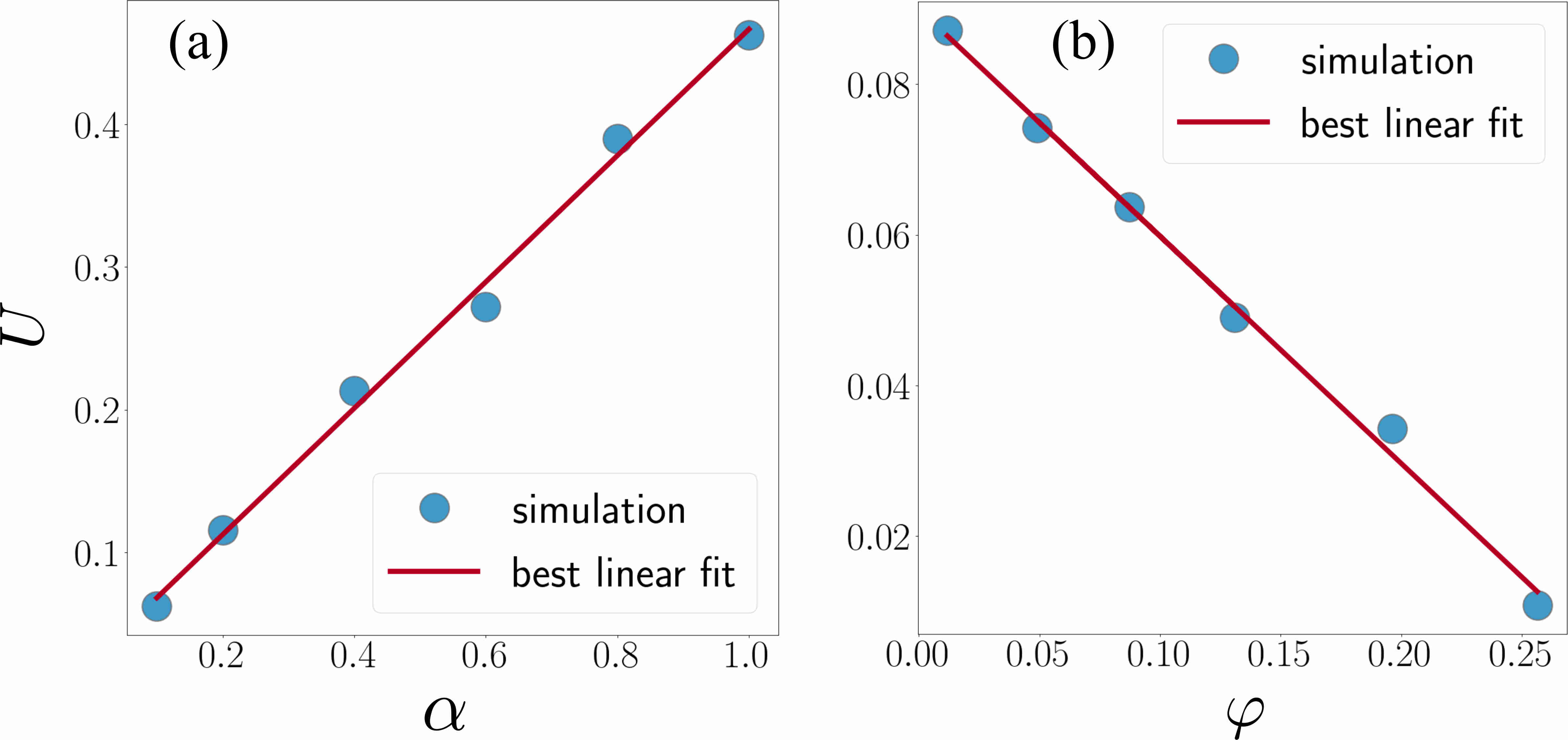} \caption{(a) The self-propulsion speed of the droplet $U$ increases linearly
with the activity parameter $\alpha$ at a fixed area fraction $\varphi=\pi R^{2}/L^{2}=0.049$.
(b) The self-propulsion speed of the droplet $U$ decreases linearly
with the area fraction $\varphi$ for a fixed $\alpha=0.2$.\
From the best linear fit of the simulation data
we obtain $U=0.44\alpha\left(1-3.4\varphi\right)$, while the theoretical
estimate of Eq.(\ref{eq:U}) is $U=0.49\alpha$ in the limit $\varphi\rightarrow 0$. 
}\label{fig:3}
\end{figure}

\emph{Self-propulsion speed, $U$:} Having described the mechanism,
we now obtain an analytical expression for $U$ using appropriate
boundary conditions corresponding to our model. 
\textcolor{black}{In simulations, the interface is diffused, 
and thus, there is no discontinuity in the fluid velocities inside ($\tilde{\boldsymbol{v}}$) and outside (${\boldsymbol{v}}$) the droplet. 
In analytical calculations, the interface is sharp,} and we solve the Stokes equation both inside and outside of the droplet using
the following boundary conditions 
\begin{subequations}\label{eq:BCatR}
\begin{align}
v_{r}(r=R) & =\tilde{v}_{r}(r=R)=0,\label{eq:no-flux}\\
v_{\theta}(r=R) & =\tilde{v}_{\theta}(r=R),\label{eq:slip}\\
\left(\boldsymbol{\sigma}-\tilde{\boldsymbol{\sigma}}\right)\cdot\hat{\boldsymbol{r}} & =\gamma\left(\boldsymbol{\nabla}\cdot\hat{\boldsymbol{r}}\right)\hat{\boldsymbol{r}}-\boldsymbol{\nabla}\gamma,\quad\text{at }r=R.\label{eq:stress}
\end{align}
\end{subequations}
The above equations represent no-flux (\ref{eq:no-flux}),
continuity of tangential slip velocity (\ref{eq:slip}) at the interface,
and the fact that the discontinuity of the Cauchy stress, $\boldsymbol{\sigma}$,
at the interface is related to the surface tension $\gamma(\theta)$
of the droplet, \emph{via} (\ref{eq:stress})~\cite{Schmitt_2016,Leven_1976}. 
The speed $U_{0}$ of
the droplet in an infinite 2D domain (see \cite{Appendix}) is then:
\begin{equation}
U_{0}=\frac{\Delta\gamma}{16\eta},\label{eq:U}
\end{equation}
As described previously $\Delta\gamma=\alpha\Delta\psi\gamma_{0}/\kappa$,
and thus, the speed increases linearly with contractile activity $\alpha$
and actomyosin concentration in the cortex $\rho$. 
Using the above formula and the parameters of Fig.~\ref{fig:3}, the theoretically predicted speed is $U_{0}=0.49\alpha$.
This is in excellent agreement with the numerical estimate of $U_{0}=0.44\alpha$ obtained from a best-fit of the simulation data of 
Fig.~\ref{fig:3}. It should be noted that the speed does not depend on $R$, which is consistent with the 
literature on transport by interfacial forces \cite{anderson1989colloid, Schmitt_2016}. 
\textcolor{black}{We ignore any deformations of the droplet in our calculations. 
A more detailed analysis studying the role of deformation will be pursued in a future work. }

To account for the periodic boundary conditions used in simulations,
we need to sum the flow due to periodic images of the droplet, which
lie on a 2D square lattice. The expression for the speed $U$ is then
$U=U_{0}(1-\varphi)$ \citep{Appendix}. The best fit of numerical
data in Fig.~\ref{fig:3} gives $U=0.44\alpha\left(1-3.4\varphi\right)$.
Although our analytical results give the linear scalings for the self-propulsion speed,
as a function of activity parameter $\alpha$ and area fraction $\varphi$,
seen in the simulations, this belies the true complexity of the problem which exhibits
linear scaling with volume fraction far beyond any perturbative regime
(and indeed with a different coefficient). A more detailed analysis
of the above will be pursed in a future work. 
\begin{figure}[t]
\includegraphics[width=0.89\columnwidth]{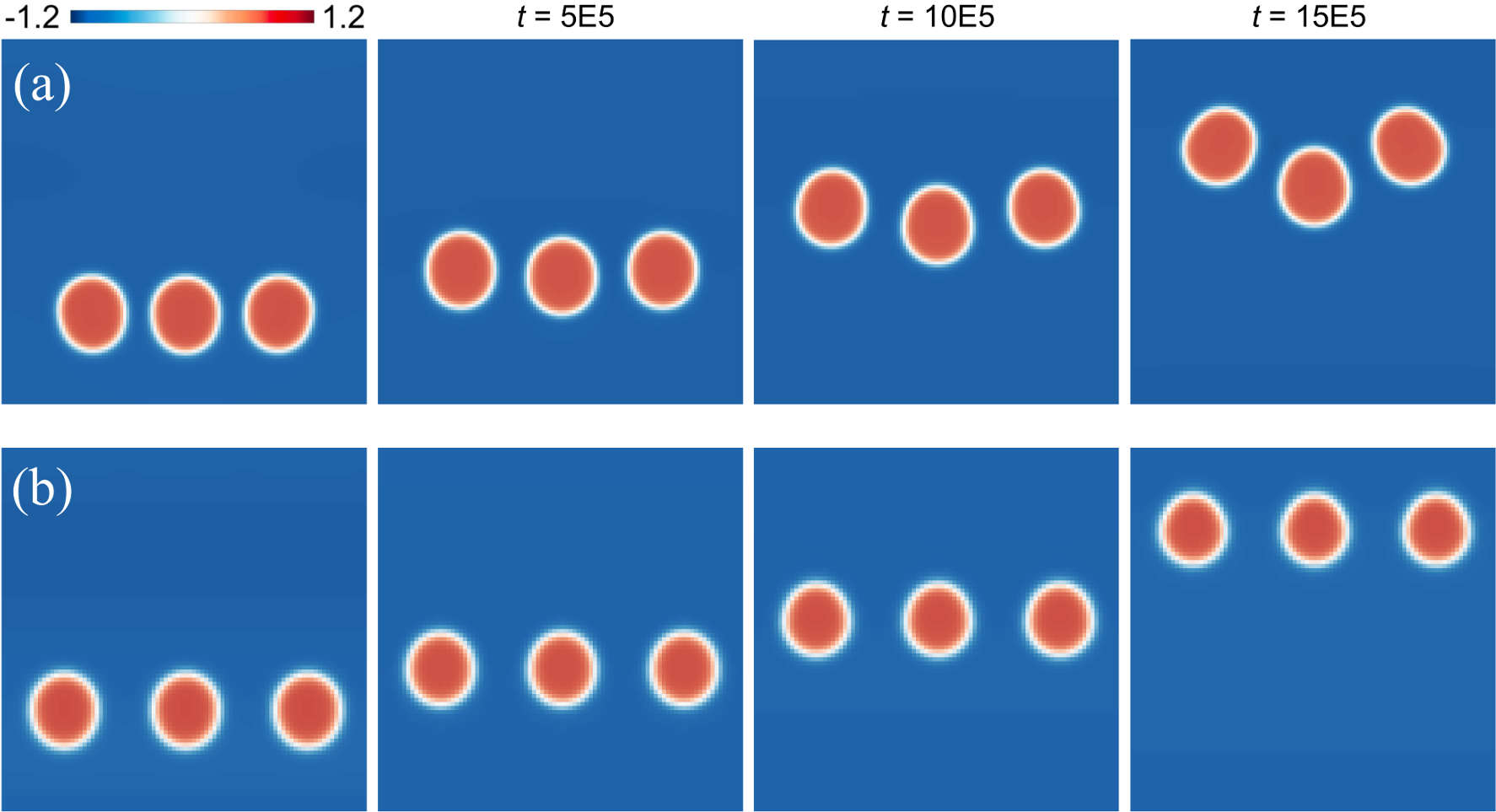} \caption{(a) Active-Crowley-like instability: A linear array of active droplets
is unstable. This instability is opposite to that in an array of sedimenting
particles. 
(b) There is no instability if
there is no well-defined central particle. 
\label{fig:4}}
\end{figure}

\emph{Crowley-like instability of an array of active droplets:} The
model and simulations above can be easily extended to the case of
multiple droplets. This is operationally done by generalizing the
free energy functional of (\ref{eq:F}) for $N$ active droplets as:
\begin{align}
\mathcal{F}[\boldsymbol{\phi},\boldsymbol{\psi}] & =\int\sum_{i=1}^{N}\Big(\frac{a}{2}\phi_{i}^{2}+\frac{b}{4}\phi_{i}^{4}+\frac{a'}{2}\psi_{i}^{2}+\frac{b'}{4}\psi_{i}^{4}-\beta\phi_{i}\psi_{i}\nonumber \\
 & +\frac{\kappa}{2}(\boldsymbol{\nabla}\phi_{i})^{2}+\frac{\kappa'}{2}(\boldsymbol{\nabla}\psi_{i})^{2}+\sum_{j\ne i}\beta'\phi_{i}\phi_{j}\Big)d\boldsymbol{r}.
\end{align}
Here the term proportional to $\beta'>0$ effectively leads to repulsion
between the droplet $\phi_{i}$ and $\phi_{j}$~\cite{Foglino_2017},
and thus, precludes any overlap. The dynamics for each $\phi_{i}$,
$\psi_{i}$ and $\boldsymbol{v}$ is the same as in (\ref{eq:phidot}-\ref{eq:Stokes})
and the equilibrium interfacial stress $\boldsymbol{\Sigma}^{E}$
and active stress $\boldsymbol{\Sigma}^{A}$ are now the sum of each
contribution from $\phi_{i}$ and $\psi_{i}$ given in (\ref{eq:SigmaE})
and (\ref{eq:SigmaA}).

We use the above to study the instability in a linear array of three
droplets as shown in Fig.~\ref{fig:4}. We see an active-Crowley-like
instability when there is a well-defined central particle. Unlike
in the case of the sedimentation~\cite{Ramaswamy_2001}, here the
central particle lags. This can be understood from the fluid flow
in Fig.~\ref{fig:2}(c), which has the effect of the central particle
being pushed backwards by the neighboring particles. On the other
hand if the particles are equally separated as in Fig.~\ref{fig:4}(b),
there is no instability as there is no well-defined central particle,
when taking into account the periodic boundary condition used.

\emph{Conclusion:} We have presented a minimal (and scalable) hydrodynamic
model of active droplets without any liquid-crystalline order parameter~\cite{Tjhung_2012,Tjhung_2015}. Nor do we require
explicit chemical reactions in the form of source and sink terms~\cite{Yabunaka_2012,Fadda_2017,Morozov_2019,laviPRE2020}.
The self-propulsion of droplets in our model relies on the 
\textcolor{black}{fact that excess contractility in the back of the droplet 
gives rise to a finite effective surface tension gradient, $\Delta\gamma \neq 0$}.
Thus, although motivated by a simple description of swimming cells, our model can also capture the self-propulsion
due to Marangoni stresses on the surface of active emulsion droplets \cite{Thutupalli_2011,Jin_2017,thutupalli2018FIPS,Izzet_2019, morozov2019self, lohse2020physicochemical}.
\textcolor{black}{The effective tension in our case is not a prescribed quantity, but is instead, a consequence of the minimal form of the active stress $\boldsymbol\Sigma^{A}$ given in Eq.(\ref{eq:SigmaA}). }
Our theory might possibly be extended to address a passive liquid crystal
inside an active scalar droplet, to mimic experimental systems of \cite{thutupalli2018FIPS,Thutupalli_2011}, or, in the chiral case, the helical trajectories seen in experiments of \cite{yamamoto2017}. 

We also showed the feasibility of our model for the study of many droplets
by studying an active Crowley-like instability in a linear array of
active droplets. A more detailed study of an active droplet suspension
using our theory and its comparison to the particulate theories \cite{shaebani2020computational}
of active matter will be presented in future work. 

RS is funded by a Royal Society-SERB Newton International Fellowship.
MEC is funded by the Royal Society. Numerical work was performed on
the Fawcett HPC system at the Centre for Mathematical Sciences. Work
funded in part by the European Research Council under the Horizon
2020 Programme, ERC grant agreement number 740269.

\vspace{2cm}

\begin{center}\textbf{\Large{}Supplemental Information (SI)}\end{center}
\section{Stokes flow of a self-propelling droplet in infinite two-dimensions
\label{app:dilute}}

In this Section, we will derive the analytic solution for the fluid
velocity in the limit of infinite boundary ($L\rightarrow\infty$)
and sharp $\phi$-interface ($\xi_{0}\rightarrow0$). We consider
a two-dimensional active droplet swimming with velocity $-U_{0}\hat{\boldsymbol{x}}$
in the lab frame. In the co-moving frame, the droplet will be stationary,
and the fluid velocity at far field $r\rightarrow\infty$ is $U_{0}\hat{\boldsymbol{x}}$.
We can then define $\boldsymbol{v}(\boldsymbol{r})$ and $\tilde{\boldsymbol{v}}(\boldsymbol{r})$
to be the fluid velocity inside and outside the droplet respectively;
$\boldsymbol{v}(\boldsymbol{r})$ and $\tilde{\boldsymbol{v}}(\boldsymbol{r})$
are obtained from solving two independent Stokes equation. We then
match the two solutions at the interface of the droplet $r=R$ \emph{via}
the boundary conditions in Eq.(6) of the main text. We can assume
the droplet to be centered at the origin. It is convenient to obtain
the solution of the fluid flow using stream-function $\vartheta$,
defined as $\boldsymbol{v}=\boldsymbol{\nabla}\times(\vartheta\hat{\boldsymbol{z}})$~\cite{Happel_1981}.
The components of the fluid velocity are given in terms of the stream-function
in Cartesian $(x,y)$ and plane polar $(r,\theta)$ coordinates as
\begin{equation}
v_{x}=\frac{\partial\vartheta}{\partial y},\quad v_{y}=-\frac{\partial\vartheta}{\partial x},\quad v_{r}=\frac{1}{r}\frac{\partial\vartheta}{\partial\theta},\quad v_{\theta}=-\frac{\partial\vartheta}{\partial r}.
\end{equation}
Now using the incompressibility condition $\boldsymbol{\nabla}\cdot\boldsymbol{v}=0$
and Stokes equation, it can be shown by standard arguments that the
stream-function satisfies the biharmonic equation: 
\begin{equation}
\nabla^{4}\vartheta=0,\label{eq:biharmonic}
\end{equation}
subject to the boundary conditions for the corresponding velocity
field given in the main text.

Now let us denote $\boldsymbol{v}=\boldsymbol{\nabla}\times(\vartheta\hat{\boldsymbol{z}})$
and $\tilde{\boldsymbol{v}}=\boldsymbol{\nabla}\times(\tilde{\vartheta}\hat{\boldsymbol{z}})$
to be the fluid velocity inside and outside the droplet respectively.

\textbf{Exterior flow $\boldsymbol{v}$:} First, we will solve the
fluid velocity outside the droplet. The boundary condition for the
fluid velocity at infinity $r\rightarrow\infty$ is: 
\begin{equation}
v_{x}=\frac{\partial\vartheta}{\partial y}=U_{0},\quad\text{as }r\rightarrow\infty.
\end{equation}
This implies that the stream-function has the following form at $r\rightarrow\infty$:
\begin{equation}
\vartheta(r,\theta)=U_{0}r\sin\theta,\quad\text{as }r\rightarrow\infty.\label{eq:BC1}
\end{equation}
Thus we can use separation of variables: 
\begin{equation}
\vartheta(r,\theta)=f(r)\sin\theta.\label{eq:separation-var}
\end{equation}
Substituting (\ref{eq:separation-var}) into (\ref{eq:biharmonic}),
$f(r)$ satisfies the following equation: 
\begin{equation}
\left(\frac{d^{2}}{dr^{2}}+\frac{1}{r}\frac{d}{dr}-\frac{1}{r^{2}}\right)^{2}f(r)=0.\label{eq:f}
\end{equation}
Using trial solution $f(r)\sim r^{n}$, we can obtain the general
solution to (\ref{eq:f}): 
\begin{equation}
f(r)=\frac{A}{r}+Br+Cr\ln r+Dr^{3}.
\end{equation}
From the boundary condition (\ref{eq:BC1}), we get $C=D=0$ and $B=U_{0}$.

Now the no-flux boundary condition for the normal component of the
velocity at the interface $r=R$ reads: 
\begin{equation}
v_{r}(R,\theta)=\frac{1}{R}\left.\frac{\partial\vartheta}{\partial r}\right|_{r=R}=0.
\end{equation}
This implies $A=-U_{0}R^{2}$. Thus the exterior fluid velocity in
polar coordinates and co-moving frame is: 
\begin{align}
v_{r}(r,\theta) & =U_{0}\left(-\frac{R^{2}}{r^{2}}+1\right)\cos\theta\label{eq:v-r}\\
v_{\theta}(r,\theta) & =-U_{0}\left(\frac{R^{2}}{r^{2}}+1\right)\sin\theta.\label{eq:v-theta}
\end{align}

\begin{figure}
\begin{centering}
\centering\includegraphics[width=0.48\textwidth]{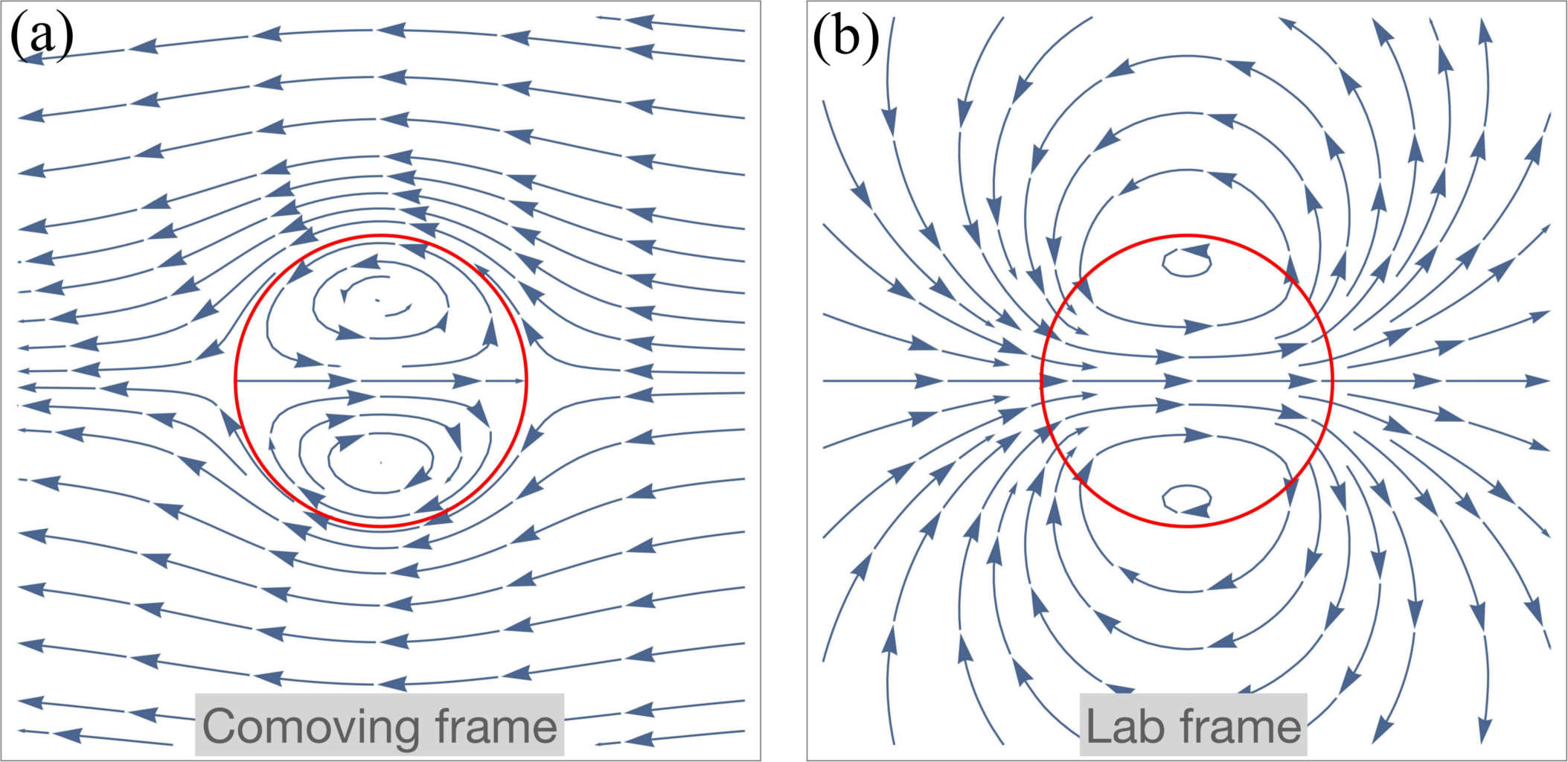} 
\par\end{centering}
\caption{Streamlines of the theoretical fluid flow around an active droplet
moving to the right in the co-moving (a) and lab-frame (b) in an infinite
domain. The red circle denotes the interface of the droplet. (Note
that the calculation in Section~\ref{app:dilute} is done with the
droplet moving to the left, thus the fluid velocity above is flipped
around the $y$-axis.) \label{fig:SI-1}}
\end{figure}

\textbf{Interior flow $\tilde{\boldsymbol{v}}$:} Again, defining
the stream-function $\tilde{\vartheta}(r,\theta)$ and using separation
of variables, the solution to the biharmonic equation (\ref{eq:biharmonic})
is 
\begin{equation}
\tilde{\vartheta}(r,\theta)=\left(\frac{\tilde{A}}{r}+\tilde{B}r+\tilde{C}r\ln r+\tilde{D}r^{3}\right)\sin\theta.
\end{equation}
Now since the fluid velocity $\tilde{\boldsymbol{v}}=\boldsymbol{\nabla}\times(\tilde{\vartheta}\hat{\boldsymbol{z}})$
has to remain finite at $r=0$, we then require $\tilde{A}=\tilde{C}=0$.
We then impose no-flux boundary condition: 
\begin{equation}
\tilde{v}_{r}(R,\theta)=\frac{1}{R}\left.\frac{\partial\vartheta}{\partial\theta}\right|_{r=R}=0,
\end{equation}
and continuity of the tangential velocity at the interface: 
\begin{equation}
\tilde{v}_{\theta}(R,\theta)=v_{\theta}(R,\theta),
\end{equation}
to finally find: $\tilde{B}=-U_{0}$ and $\tilde{D}=U_{0}/R^{2}$.
Thus the interior fluid velocity in polar coordinates is: 
\begin{align}
\tilde{v}_{r}(r,\theta) & =U_{0}\left(-1+\frac{r^{2}}{R^{2}}\right)\cos\theta\\
\tilde{v}_{\theta}(r,\theta) & =U_{0}\left(1-\frac{3r^{2}}{R^{2}}\right)\sin\theta\label{eq:vtilde-theta}
\end{align}
This fluid flow, derived from above expressions for the interior and
exterior region, has been plotted in Fig.(\ref{fig:SI-1}). This can
be compared with direct numerical simulations in the main text. It
is worthwhile to note that there is no solution for Stokes flow around
a disc, the so-called Stokes paradox \cite{Happel_1981}, if a no-slip
boundary condition at $r=R$ is used. The solution described above
relies on the fact that there is a free-slip boundary condition on
the surface of the droplet.

\textbf{Self-propulsion speed:} The hydrodynamic stress is discontinuous
at the interface $r=R$ due to the surface tension $\gamma(\theta)$:
\begin{equation}
\left(\boldsymbol{\sigma}-\tilde{\boldsymbol{\sigma}}\right)\cdot\hat{\boldsymbol{r}}=\gamma\left(\boldsymbol{\nabla}\cdot\hat{\boldsymbol{r}}\right)\hat{\boldsymbol{r}}-\boldsymbol{\nabla}\gamma,\quad\text{at }r=R.\label{eq:stress-condition}
\end{equation}
The stress in polar coordinates is: 
\begin{align}
\sigma_{rr} & =-p+2\eta\frac{\partial v_{r}}{\partial r}\\
\sigma_{\theta\theta} & =-p+2\eta\left(\frac{1}{r}\frac{\partial v_{\theta}}{\partial\theta}+\frac{v_{r}}{r}\right)\\
\sigma_{\theta r} & =\sigma_{r\theta}=\eta\left(\frac{1}{r}\frac{\partial v_{r}}{\partial\theta}+\frac{\partial v_{\theta}}{\partial r}-\frac{v_{\theta}}{r}\right).
\end{align}

First, let us look at the $r$-component of (\ref{eq:stress-condition}):
\begin{equation}
\sigma_{rr}-\tilde{\sigma}_{rr}=\frac{\gamma}{R}\quad\Longrightarrow\quad\tilde{p}-p=\frac{\gamma}{R},
\end{equation}
which is just the Laplace pressure difference between the interior
and exterior of the droplet. Now we look at the $\theta$-component
of (\ref{eq:stress-condition}): 
\begin{align}
\sigma_{\theta r}-\tilde{\sigma}_{\theta r} & =-\frac{1}{R}\frac{d\gamma}{d\theta}\\
\eta\frac{\partial v_{\theta}}{\partial r}-\tilde{\eta}\frac{\partial\tilde{v}_{\theta}}{\partial r} & =-\frac{1}{R}\frac{d\gamma}{d\theta},\quad\text{at }r=R\label{eq:stress-theta}
\end{align}
where $\eta$ and $\tilde{\eta}$ is the fluid viscosity inside and
outside the droplet respectively. Now substituting (\ref{eq:v-theta})
and (\ref{eq:vtilde-theta}) into (\ref{eq:stress-theta}), we get:
\begin{equation}
U_{0}\sin\theta=-\frac{1}{8\eta}\frac{d\gamma}{d\theta}.\label{eq:speed}
\end{equation}
Thus, the velocity of the droplet is non-zero only if we have a surface
tension gradient along the tangential direction. Now we can average
(\ref{eq:speed}) over the angle $\theta\in[0,\pi]$ to obtain: 
\begin{equation}
U_{0}=\frac{\Delta\gamma}{16\eta},\label{eq:speed2}
\end{equation}
where $\Delta\gamma$ is the surface tension difference between the
back and the front. Note that $\theta=\pi$ corresponds to the front
the droplet and $\theta=0$ corresponds to the back of the droplet.
For active model H~\cite{Tiribocchi_2015}, the effective surface
tension is $\gamma=\gamma_{0}(1-\alpha\psi/\kappa)$, and thus we
get: 
\begin{equation}
U_{0}=\frac{\gamma_{0}\alpha}{16\eta\kappa},\label{eq:speed3}
\end{equation}
since $\psi=0$ in the front and $\psi=-1$ in the back (see Fig.~1
in the main text). Thus, we obtain linear scaling with activity $\alpha$,
consistent with the numerical result in the main text.


\section{Active droplets on a lattice \label{app:lattice}}

To compute this finite size scaling theoretically, we consider an
infinite array of active droplets, whose centre of mass are located
on a square lattice. The renormalized fluid velocity in the region
around the central droplet, located at the origin, is the sum of all
the fluid velocities generated by each droplet on the lattice. The
numerical simulations presented in the main text assume periodic boundary
condition on each side of the box: $(x,\pm L/2),(\pm L/2,y)$. This
is equivalent to having an infinite number of active droplets located
on a square lattice $(mL,nL)$, where $m,n\in\mathbb{Z}$. In this
Section, we will derive the approximate fluid velocity generated by
these droplets.

First, let us consider the dilute limit $R/L\rightarrow0$ (Section~\ref{app:dilute}).
Let us consider an active droplet swimming with velocity $U_{0}\hat{\boldsymbol{x}}$
and located at the origin. The fluid flow generated by this single
droplet in the lab frame is (see Fig.~\ref{fig:SI-1}(b) and Eqns.~(\ref{eq:v-r}-\ref{eq:v-theta})):
\begin{align}
u_{x} & =U_{0}R^{2}\frac{x^{2}-y^{2}}{(x^{2}+y^{2})^{2}}\\
u_{y} & =2U_{0}R^{2}\frac{xy}{(x^{2}+y^{2})^{2}}.
\end{align}
In particular, the fluid flow at the leading edge of a single droplet
$(x,y)=(R,0)$ is equal to the velocity of the droplet itself: 
\begin{equation}
u_{x}(R,0)=U_{0}.\label{eq:ux-R-dilute}
\end{equation}

\begin{figure}
\begin{centering}
\centering\includegraphics[width=0.35\textwidth]{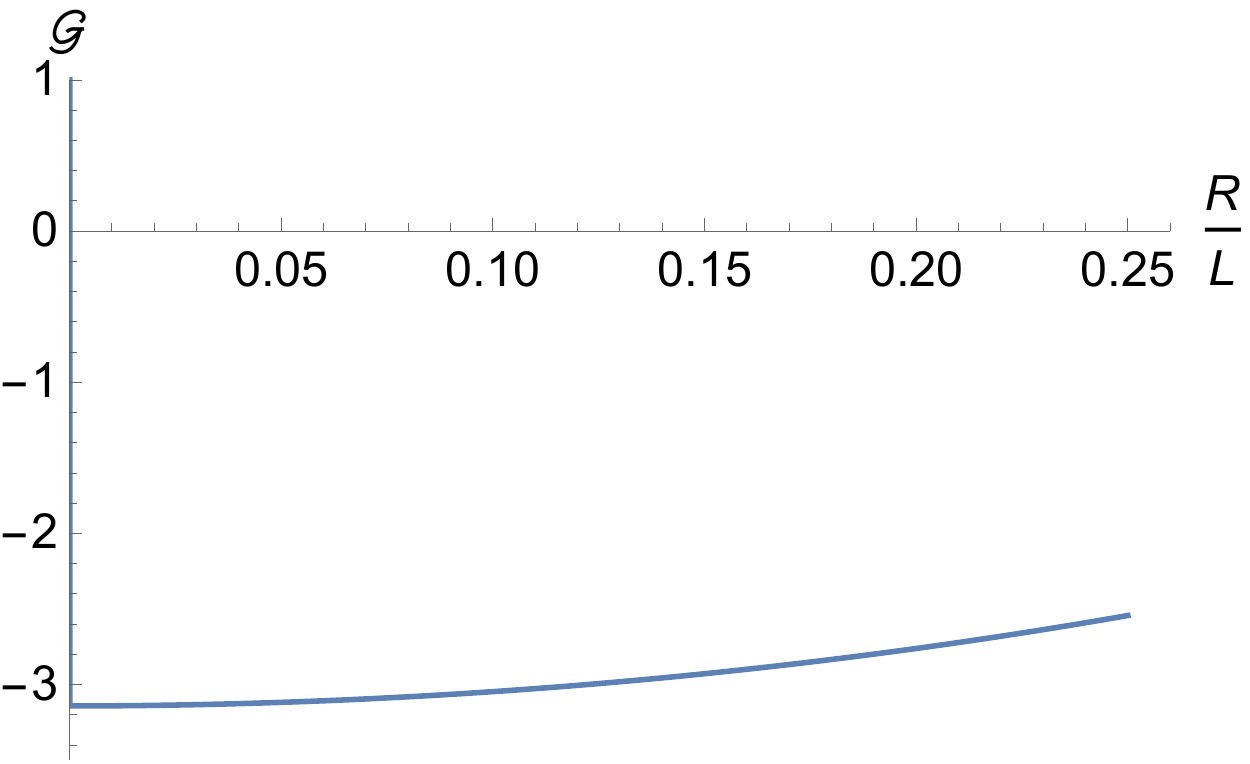} 
\par\end{centering}
\caption{Shows the plot of the infinite sum in (\ref{eq:ux-R-1}) as a function
of $R/L$, \label{fig:SI-2}}
\end{figure}

Now, the fluid flow generated by infinite droplets in a lattice is
given by (assuming dilute flow solution for each droplet): 
\begin{align}
u_{x}= & U_{0}R^{2}\frac{x^{2}-y^{2}}{(x^{2}+y^{2})^{2}}\nonumber \\
 & +U_{0}R^{2}\sum_{m=1}^{\infty}\sum_{n=0}^{\infty}\frac{(x-mL)^{2}-(y-nL)^{2}}{\left[(x-mL)^{2}+(y-nL)^{2}\right]^{2}}\nonumber \\
 & +U_{0}R^{2}\sum_{m=0}^{-\infty}\sum_{n=1}^{\infty}\frac{(x-mL)^{2}-(y-nL)^{2}}{\left[(x-mL)^{2}+(y-nL)^{2}\right]^{2}}\nonumber \\
 & +U_{0}R^{2}\sum_{m=0}^{\infty}\sum_{n=-1}^{-\infty}\frac{(x-mL)^{2}-(y-nL)^{2}}{\left[(x-mL)^{2}+(y-nL)^{2}\right]^{2}}\nonumber \\
 & +U_{0}R^{2}\sum_{m=-1}^{-\infty}\sum_{n=0}^{-\infty}\frac{(x-mL)^{2}-(y-nL)^{2}}{\left[(x-mL)^{2}+(y-nL)^{2}\right]^{2}}\ 
\end{align}
The first term is the fluid flow created by the active droplet at
the origin. In particular, the fluid flow at the edge of the centre
droplet is given by: 
\begin{align}
u_{x}(R,0) & =U_{0}\nonumber \\
 & +U_{0}\frac{R^{2}}{L^{2}}\sum_{m=1}^{\infty}\sum_{n=0}^{\infty}\frac{\left(\frac{R}{L}-m\right)^{2}-n^{2}}{\left[\left(\frac{R}{L}-m\right)^{2}+n^{2}\right]^{2}}\nonumber \\
 & +U_{0}\frac{R^{2}}{L^{2}}\sum_{m=0}^{\infty}\sum_{n=1}^{\infty}\frac{\left(\frac{R}{L}+m\right)^{2}-n^{2}}{\left[\left(\frac{R}{L}+m\right)^{2}+n^{2}\right]^{2}}\nonumber \\
 & +U_{0}\frac{R^{2}}{L^{2}}\sum_{m=0}^{\infty}\sum_{n=1}^{\infty}\frac{\left(\frac{R}{L}-m\right)^{2}-n^{2}}{\left[\left(\frac{R}{L}-m\right)^{2}+n^{2}\right]^{2}}\nonumber \\
 & +U_{0}\frac{R^{2}}{L^{2}}\sum_{m=1}^{\infty}\sum_{n=0}^{\infty}\frac{\left(\frac{R}{L}+m\right)^{2}-n^{2}}{\left[\left(\frac{R}{L}+m\right)^{2}+n^{2}\right]^{2}}\label{eq:ux-R-1}\\
 & \equiv U_{0}\left[1+\frac{R^{2}}{L^{2}}\mathcal{G}\left(\frac{R}{L}\right)\right],\label{eq:ux-R-2}
\end{align}
where $\mathcal{G}(R/L)$ is the infinite sum contained in (\ref{eq:ux-R-1}).
It can be shown that the infinite sum converges and $\mathcal{G}(R/L)$
as a function of $R/L$ is shown in the plot of Fig.~\ref{fig:SI-2}.
From the plot, $\mathcal{G}(R/L)\simeq$ constant for small $R/L$
and the constant value is $-\pi$. Therefore, the fluid velocity at
$(x,y)=(R,0)$ is 
\begin{equation}
u_{x}(R,0)=U_{0}\left[1-\pi\frac{R^{2}}{L^{2}}+\mathcal{O}\left(\frac{R^{4}}{L^{4}}\right)\right].\label{eq:ux-R-3}
\end{equation}
Now we can compare (\ref{eq:ux-R-3}) to (\ref{eq:ux-R-dilute}) to
deduce that the velocity of the central droplet at the origin is renormalized
by the neighboring droplets in the lattice. The fluid velocity of
the droplets in a square lattice as a function of area fraction $\varphi=\pi R^{2}/L^{2}$
is then: 
\begin{equation}
U\simeq U_{0}\left(1-\varphi\right)=\frac{\Delta\gamma}{16\eta}\left(1-\varphi\right).
\end{equation}
Thus we find linear scaling with decreasing $\varphi$ and increasing
$\alpha$ or $\Delta\gamma$, consistent with the numerical result
in the main text, although the pre-factors are not exactly matched.
The reason is because the neighboring droplets also renormalize the
fluid velocity at the interface of the central droplet $r=R$. Thus
the stress boundary condition (\ref{eq:stress-condition}), no-flux
condition, and continuity of tangential velocity at the interface
has to be reevaluated. Secondly in simulations, inside the $\phi$-droplet,
there is also a $\psi$-droplet, which gets squashed. This contributes
to a dipolar fluid flow \cite{Singh_2019}, while we only consider
the quadrupolar flow in our analysis. A more detailed study, accounting
for all these effects and the role of periodic images in evaluation
the fluid flow at all orders using a absolutely convergent expression
of the fluid flow \cite{brady1988dynamic,o1979method}, will be presented
in a future work.

\section{Simulation details}

The fluid flow satisfies the Stokes equation 
\begin{align}
0 & =-\boldsymbol{\nabla}p+\eta\nabla^{2}\boldsymbol{v}+\boldsymbol{f}\label{eq:stokes1}\\
0 & =\boldsymbol{\nabla}\cdot\boldsymbol{v},
\end{align}
where, 
\begin{equation}
\boldsymbol{f}=\boldsymbol{\nabla}\cdot(\boldsymbol{\Sigma}^{A}+\boldsymbol{\Sigma}^{E}),
\end{equation}
is the force density on the fluid. The solution is obtained using
Fourier transforms as we now describe.

Defining the Fourier transform of a function $\phi(\boldsymbol{r})$
as 
\begin{align}
\phi_{\boldsymbol{k}} & =\int\phi(\boldsymbol{r})e^{-i\boldsymbol{k}\cdot\boldsymbol{r}}\,d\boldsymbol{r}\\
\phi(\boldsymbol{r}) & =\frac{1}{(2\pi)^{3}}\int\phi_{\boldsymbol{k}}e^{i\boldsymbol{k}\cdot\boldsymbol{r}}\,d\boldsymbol{k},
\end{align}
we can write the Stokes equation (\ref{eq:stokes1}) in the Fourier
space as 
\begin{align}
0 & =-i\boldsymbol{k}p_{\boldsymbol{k}}-\eta k^{2}\boldsymbol{v}_{\boldsymbol{k}}+\boldsymbol{f}_{\boldsymbol{k}},\label{eq:stokes-k}\\
0 & =i\boldsymbol{k}\cdot\boldsymbol{v}_{\boldsymbol{k}}.\label{eq:incompressibility}
\end{align}
The above equations can be used to project out the solution of the
pressure field in Fourier space: 
\begin{equation}
p_{\boldsymbol{k}}=-i\frac{\boldsymbol{k}\cdot\boldsymbol{f}_{\boldsymbol{k}}}{k^{2}}.
\end{equation}
The above expression of the pressure is then used in (\ref{eq:stokes-k})
to obtain the solution of the fluid flow, with built-in incompressibility,
given as 
\begin{align}
\boldsymbol{v}_{\boldsymbol{k}} & =\mathbf{G}_{\boldsymbol{k}}\cdot\boldsymbol{f}_{\boldsymbol{k}}\\
\mathbf{G}_{\boldsymbol{k}} & =\frac{1}{\eta}\left(\frac{\boldsymbol{I}}{k^{2}}-\frac{\boldsymbol{k}\boldsymbol{k}}{k^{4}}\right).\label{eq:stokes-k-1}
\end{align}
Here $\mathbf{G}_{\boldsymbol{k}}$ is the Fourier transform of the
Oseen tensor $\mathbf{G}(\boldsymbol{r})$ \cite{Pozrikidis_1992}.

The above solution for the fluid flow is implemented using the standard
fast Fourier transforms (FFTs) implemented in NumPy \cite{oliphant_2006},
which automatically implements periodic boundary condition. The remaining
terms in the dynamics of the $\phi$ and $\psi$ field are implemented
using using the pseudo-spectral method, involving again NumPy Fourier
transforms and $2/3$ dealiasing procedure \cite{Orszag_1971,Boyd_2001}.
The linear terms are directly evaluated in the Fourier space, while
the non-linear terms are computed in real space by inverse Fourier
transforms. This is then transformed back to Fourier space to evolve
the dynamical system in time using the explicit Euler-Maruyama method
\cite{Kloeden_1992}. We provide the parameters used in generating
all the figures of the manuscript in Table \ref{tab:Table-of-simulation-parameters}.
Finally, we explain the method used to determine the speed and centre-of-mass
position of the $\phi$-droplet. Using the simulation data for the
field $\phi(\boldsymbol{r},t)$, we obtain the centre-of-mass coordinate
$\boldsymbol{R}^{\text{CM}}$ as 
\begin{equation}
\boldsymbol{R}^{\text{CM}}(t)=\frac{\int\boldsymbol{r}\phi^{s}(\boldsymbol{r},t)\,d\boldsymbol{r}}{\int\phi^{s}(\boldsymbol{r},t)\,d\boldsymbol{r}},
\end{equation}
where $\phi^{s}(\boldsymbol{r},t)$ is defined to be $1$ if $\phi(\boldsymbol{r},t)\ge0$
and $0$ if $\phi(\boldsymbol{r},t)<0$. The self-propulsion velocity
of the droplet is obtained using: 
\begin{align}
\boldsymbol{U}(t) & =\frac{\int\phi^{s}(\boldsymbol{r},t)\boldsymbol{v}(\boldsymbol{r})\,d\boldsymbol{r}}{\int\phi^{s}(\boldsymbol{r},t)\,d\boldsymbol{r}}.
\end{align}
The above expression for the droplet velocity can be verified to hold
by using the definition $\boldsymbol{U}(t)=d\boldsymbol{R}^{\text{CM}}(t)/dt$
and Eqs.~(1-3) of the main text. 
\begin{table*}
\global\long\def\arraystretch{2.2}%
\begin{tabular}{|>{\centering}b{7cm}|>{\centering}p{6cm}|}
\hline 
Physical parameters & Model parameters\tabularnewline
\hline 
\hline 
Passive surface tension ($\gamma_{0}$) \cite{bray1994theory,Cates_2017} & $\gamma_{0}\simeq\sqrt{-8\kappa a^{3}/9b^{2}}$\tabularnewline
\hline 
Effective surface tension ($\gamma$) in presence of active stress
\cite{Cates_2017,Singh_2019,Tiribocchi_2015} & $\gamma=\gamma_{0}(1-\alpha\psi/\kappa)$\tabularnewline
\hline 
Ratio of active and passive surface tensions
& $\gamma/\gamma_0=1-\alpha\psi/\kappa$\tabularnewline
\hline 
Binodal density of droplet $\phi$ \cite{bray1994theory,Cates_2017} & $\phi_{b}=\pm\sqrt{-a/b}$\tabularnewline
\hline 
Interfacial width of droplet $\phi$ & $\xi_{\phi}\simeq\sqrt{-2\kappa/a}$\tabularnewline
\hline 
Binodal density of droplet $\psi$ \cite{bray1994theory,Cates_2017} & $\psi_{b}=\pm\sqrt{-a'/b'}$\tabularnewline
\hline 
Interfacial width of droplet $\psi$ & $\xi_{\psi}\simeq\sqrt{-2\kappa'/a'}$\tabularnewline
\hline 
Radius of the $\phi$-droplet & $R$\tabularnewline
\hline 
Radius of the $\psi$-droplet & $0.9 R$\tabularnewline
\hline 
Average thickness of the cortex ($\rho=\phi-\psi$) & $0.1 R$\tabularnewline
\hline 
Reynolds number & $\text{Re}=\frac{\text{Inertial force}}{\text{Viscous force}}$\tabularnewline
\hline 
Capillary number & ${\displaystyle \mathrm{Ca}=\eta U_{0}/\gamma_{0}}$\tabularnewline
\hline 
\end{tabular}\caption{\label{tab:parametersPhysicalModel}Physical parameters and their
corresponding model parameters used in this paper. See Tab:\ref{tab:Table-of-simulation-parameters}
for values of the parameters. In this paper, we set inertial to zero,
and thus, $\text{Re}=0$, while typical $\mathrm{Ca}\sim0.001-0.01$. }
\end{table*}
\begin{table*}
\global\long\def\arraystretch{2.2}%
\begin{tabular}{|>{\centering}b{2cm}|>{\centering}p{3cm}|>{\centering}p{2cm}|>{\centering}p{2cm}|>{\centering}p{2cm}|>{\centering}p{2cm}|}
\hline 
Figure  & System size ($L\times L$)  & $R$  & $\alpha$  & $\beta$  & $\beta'$\tabularnewline
\hline 
\hline 
2 (a)  & $100\times100$  & 20  & 0.1  & 0.02  & 0\tabularnewline
\hline 
2 (b)  & $100\times100$  & 20  & 0.1  & 0.02  & 0\tabularnewline
\hline 
2 (c)  & $100\times100$  & 20  & 0.1  & 0.02  & 0\tabularnewline
\hline 
3 (a)  & $100\times100$  & 12.5  & varied  & 0.02  & 0\tabularnewline
\hline 
3 (b)  & $100\times100$  & varied  & 0.2  & 0.02  & 0\tabularnewline
\hline 
4 (a)  & $120\times120$  & 10  & 0.2  & 0.02  & 0.2\tabularnewline
\hline 
4 (b)  & $90\times90$  & 10  & 0.2  & 0.02  & 0.2\tabularnewline
\hline 
\end{tabular}\caption{\label{tab:Table-of-simulation-parameters}Simulation parameters used
to study the self-propulsion of active droplets. Throughout the paper,
the parameters of the free energy are fixed to be: $a=-b=a'=-b'=-1$,
$\kappa=1.45$, and $\kappa'=3$ for one droplet simulation. The simulations
reported are in two space dimensions, while the analytic predictions
would be maintained in higher dimensions. The parameter $\beta$ is
fixed to be small and positive $\beta=0.02$ so that the $\psi$-droplet
is confined inside the larger $\phi$-droplet. We keep the ratio of
the radius of the two droplets fixed at $0.9$, such that $\phi$
droplet is bigger than the $\psi$ droplet. We fix the mobilities
to be unity: $M^{\phi}=M^{\psi}=1$ and the viscosity to be $\eta=0.1$.
The values of $L$, $R$, and $\alpha$ vary. We also define the area
fraction to be $\varphi=\pi R^{2}/L^{2}$ . The time discretization
is fixed to be $\Delta t=0.001$ and spatial discretization is fixed
to be $\Delta x=1$. }
\end{table*}

\section{Role of dipolar flow}
\begin{figure}
\begin{centering}
\centering\includegraphics[width=0.35\textwidth]{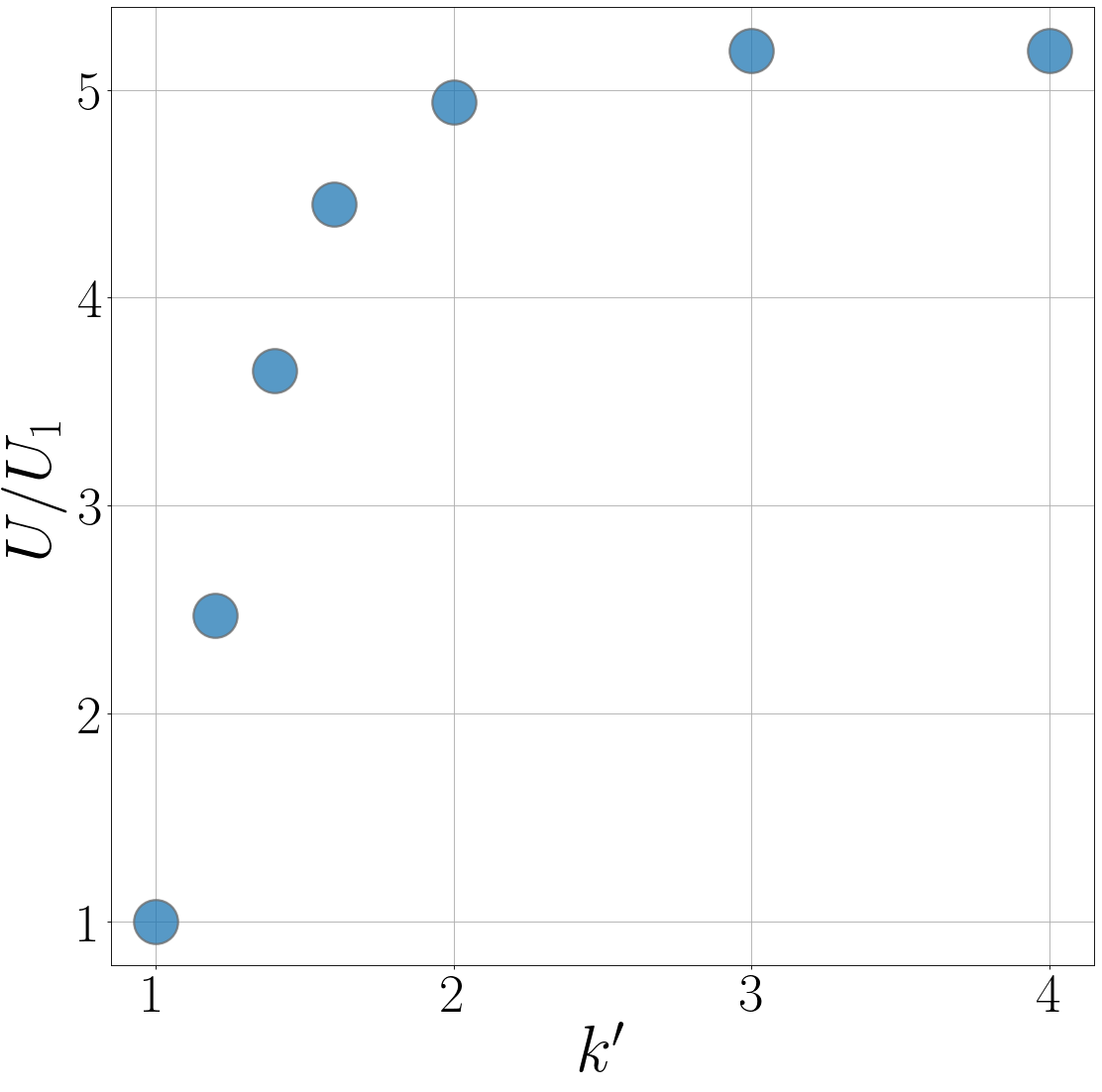} 
\par\end{centering}
\caption{Speed $U$ of a droplets as the parameter $\kappa'$ is varied. The
speed is scaled by $U_{1}$, the speed for $\kappa'=1$. \label{fig:SI-3}}
\end{figure}
The passive stress of Eq.(4) in the main text leads to a dipolar fluid
flow in presence of deformations in the $\phi$ and $\psi$ field.
The deformations, in turn, are controlled by the stiffness $\kappa$
and $\kappa'$. We use $\kappa=1.45$ to reduce deformations in the
$\phi$ field. We then study the effect of changing the parameter
$\kappa'$. The dipolar flow due to deformation of the $\psi$ field
impeded the droplets. In Fig.(\ref{fig:SI-3}) we show that the speed 
increases as we increase the parameter $k'$ and saturates. We choose 
$\kappa'=3$ when the speed has saturated. 

\section{Supplemental movies}

The two supplemental movies are: 
\begin{itemize}
\item 
{Movie
I: }The movie corresponds to results shown in Fig.2 of the main text
with the same set of parameters.

\item {Movie II: }The movie shows dynamics of two droplets simulated
using Eq.(9) of the main text. Parameters are same as in Fig.(4) but
for two droplets.
\end{itemize}

\end{document}